\documentclass[runningheads]{svmult}
\usepackage{epsfig}
\usepackage{makeidx}   
\usepackage{graphicx}  
\usepackage{subeqnar}  
\usepackage{multicol}  
\usepackage{physmubb}  
\makeindex             

\newcommand{\PO}{I\!\!P}
\newcommand{\RO}{I\!\!R}
\newcommand{\xpom}{x_{\PO}}

\begin{document}

\title*{Selected Topics in Rapidity Gap Physics}
\titlerunning{Rapidity Gap Physics}
\toctitle{Selected topics in Rapidity Gap Physics}

\author{{\large Jeffrey~R.~Forshaw} \\ Department of Physics \& Astronomy \\
University of Manchester. \\ Manchester. M13 9PL. \\ UK.}
\authorrunning{Jeffrey~R.~Forshaw}

\maketitle              

\vspace*{-6cm}
  \begin{flushright}
    MC-TH-2002/13 \\
  \end{flushright}

\vspace*{6cm}

\begin{abstract}
This talk\footnote{Talk presented at the 14th
Topical Conference on Hadron Collider Physics, HCP 2002, Karlsruhe,
29 September--4 October 2002.}
will review selected topics in rapidity gap physics. In particular
I will discuss diffractive jet production and the possibility of searching
for the higgs boson using diffraction at the LHC; the dipole picture of
diffraction and saturation; and those processes where a large momentum is
transferred across the rapidity gap, for which there has been recent
progress both experimentally and theoretically. 

\end{abstract}

\section{Introduction}
Over the past 10 years, due in no small part to the quality and extent of
data collected at the HERA and Tevatron colliders, the field of rapidity gap 
physics has flourished. As a result, in a review talk like this I cannot 
hope to cover anything other than a few topics, chosen to reflect my 
personal bias. 

The next section
will focus on the hard diffractive production of jets and higgs bosons, and
will draw on data collected at both HERA and the Tevatron. In Section 3, I
discuss the dipole model of diffraction and the evidence for saturation.
In Section 4, I turn to rapidity gaps with a large momentum transfer across
the gap. These are rarer but rather clean processes, and recent high quality
data on vector meson production has allowed comparison with
theory, which I discuss.

\section{Hard diffraction}
\begin{figure}
\begin{center}
\includegraphics[width=.6\textwidth]{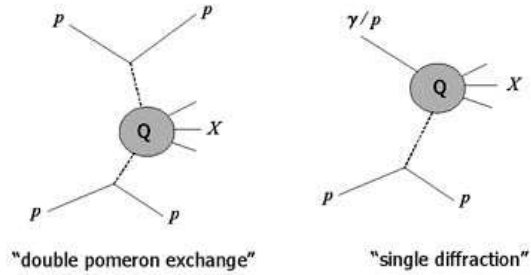}
\end{center}
\caption[]{Hard diffraction}
\label{fig:diffrac}
\end{figure}
I will follow the conventional language and classify hard
diffractive processes as shown in Figure \ref{fig:diffrac}. $Q$ is some
hard scale (e.g. jet transverse momentum, $W$ mass etc.) characteristic of
the system $X$. In ``double pomeron exchange'', which has been measured
at the Tevatron \cite{CDF:DPE}, 
the protons remain intact, losing only a small fraction
of their initial energy, so that the system $X$ is produced centrally. In
single diffraction, only one proton remains intact and fast, and the system 
$X$ is distant from it in rapidity.

\begin{figure}
\begin{center}
\includegraphics[width=.7\textwidth]{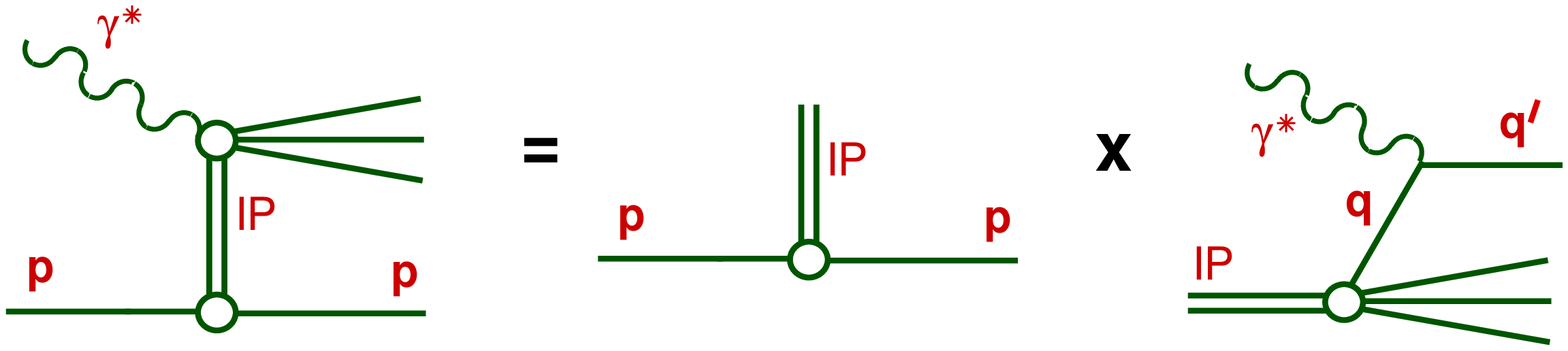}
\end{center}
\caption[]{Regge factorization. (Figure from \cite{Thompson})}
\label{fig:regge}
\end{figure}

\begin{figure}
\begin{center}
\includegraphics[width=.65\textwidth]{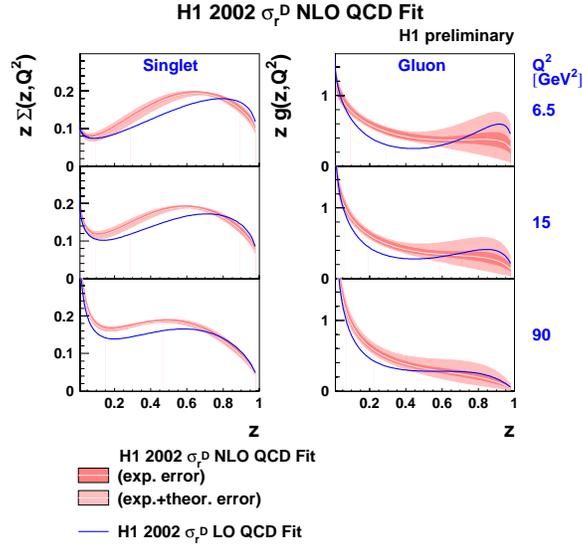}
\end{center}
\caption[]{Pomeron parton density functions extracted by H1 \cite{Schilling}}
\label{fig:h1partons}
\end{figure}

At HERA, single diffraction of a virtual photon (virtuality $Q^2$) has allowed
experimenters to probe the partonic structure of the diffractive exchange
\cite{H1:DDIS,ZEUS:DDIS}.
Appealing to regge theory, one can attempt to write the cross-section as
a product of a pomeron flux factor $f_{\PO/p}$ and a pomeron parton
density function, as shown in Figure \ref{fig:regge}. The experimenters
parameterise the flux factor as
\begin{equation}
f_{\PO/p}(\xpom,t) = \frac{e^{Bt}}{\xpom^{2 \alpha(t)-1}}
\end{equation}
where $\xpom$ is the fraction of the incoming proton's energy carried by
the pomeron, $t$ is the momentum transfer to the scattered proton, $B$ is 
the diffractive slope and $\alpha(t)$ is the pomeron trajectory. The 
experimenters are able to fit all their data on the diffractive structure
function using a parameterisation of this form after evolving the parton
density functions using the NLO evolution equations. H1 finds a pomeron
intercept $\alpha_{\PO} = 1.17 \pm 0.02$ and no need for secondary regge
exchanges for $\xpom < 0.01$ \cite{Schilling}. 
H1 has now extracted the pomeron quark and
gluon density functions with an estimate of the error, see Fig. 
\ref{fig:h1partons}. Note that it is the gluon at large $z=x/\xpom$ which
is least well constrained.

If the notion of a universal pomeron parton density function is to be
tested, one needs to take the partons as measured in diffractive DIS, and
use them to predict the rates for other processes. At HERA this programme
has been carried out quite extensively and on the whole there is good 
agreement \cite{Thompson}. However, there are indications that the DIS partons
do tend to lead to an overestimate of the data collected off real photons.
Good agreement can be arranged if one is prepared to accept an overall
renormalisation by a factor of about $0.6$ \cite{Schaetzel}. 
Strictly
speaking, the need for such a renormalisation violates universality. This
should not come as a surprise; we already knew that universality should
not hold across the board in diffraction \cite{facbreak}. 
What would be of interest is if
the violation can be understood. Simple (eikonal) models predict that
rapidity gaps will be filled in by secondary interactions in those processes
where the incoming beam particles have structure, which is the case in
hadron-hadron interactions and photon-hadron interactions with
an on-shell photon. These simple models also predict that the filling in of
gaps can be approximated by an overall multiplicative factor which is
weakly process dependent; depending primarily on the overall 
centre-of-mass energy \cite{gapsurvival}. In this way we can understand a 
gap survival factor of $\sim 0.6$ at HERA. 
The eikonal models also predict a survival factor
of around $0.1$ at the Tevatron (the high centre-of-mass energy being the 
main reason for the reduction since it liberates more low-$x$ partons). 

The burning question is therefore: ``How does a gap survival factor of 0.1, in
conjunction with the latest H1 parton density functions, stand up to the 
Tevatron data?''. At first sight, the answer is ``very badly''. In Table 
\ref{tab:teva}, we show the original calculations of \cite{Whitmore}, which are
parton level and do not contain any gap survival factor. 
Comparison is to
CDF and D0 data available at the time (some of which were preliminary). The
references shown in the table are to the final published papers. The pomeron
parton densities do not now agree with the most recent HERA data, but they
are not so far out to account for the obvious problems. Even with a
gap survival of $\sim 0.1$, things look bleak.  

\begin{table}
\caption{Problems at the Tevatron?}
\begin{center}
\renewcommand{\arraystretch}{1.4}
\setlength\tabcolsep{5pt}
\begin{tabular}{lcrr}
\hline\noalign{\smallskip}
  & Experiment & Theory  & Theory/Exp.\\
\noalign{\smallskip}
\hline
\noalign{\smallskip}
CDF W \cite{CDF:W} & $1.15 \pm 0.51 \pm 0.20 \%$ & $7\%$ & $6 \pm 3$ \\
CDF RG dijet \cite{CDF:RJJ} 
& $0.75 \pm 0.05 \pm 0.09 \%$ & $16\%$ & $22 \pm 3$ \\
CDF pot dijet \cite{CDF:FDJJ} & $0.109 \pm 0.003 \pm 0.016 \%$ & $4 \%$ & $34 \pm 5$ \\
D0 RG dijet \cite{D0:dijets}& $0.67 \pm 0.05 \%$ & $12 \%$ & $18 \pm 1$ \\
CDF heavy quark \cite{CDF:HQ} & $0.18 \pm 0.03 \%$ & $30 \%$ & $167 \pm 28$ \\
CDF double pomeron exch. \cite{CDF:DPE} &  $13.6 \pm 2.8 \pm 2$ nb & 3713 nb & $273 \pm 69$ \\
\hline
\end{tabular}
\end{center}
\label{tab:teva}
\end{table}

\begin{figure}
\begin{center}
\includegraphics[clip=true,width=.55\textwidth]{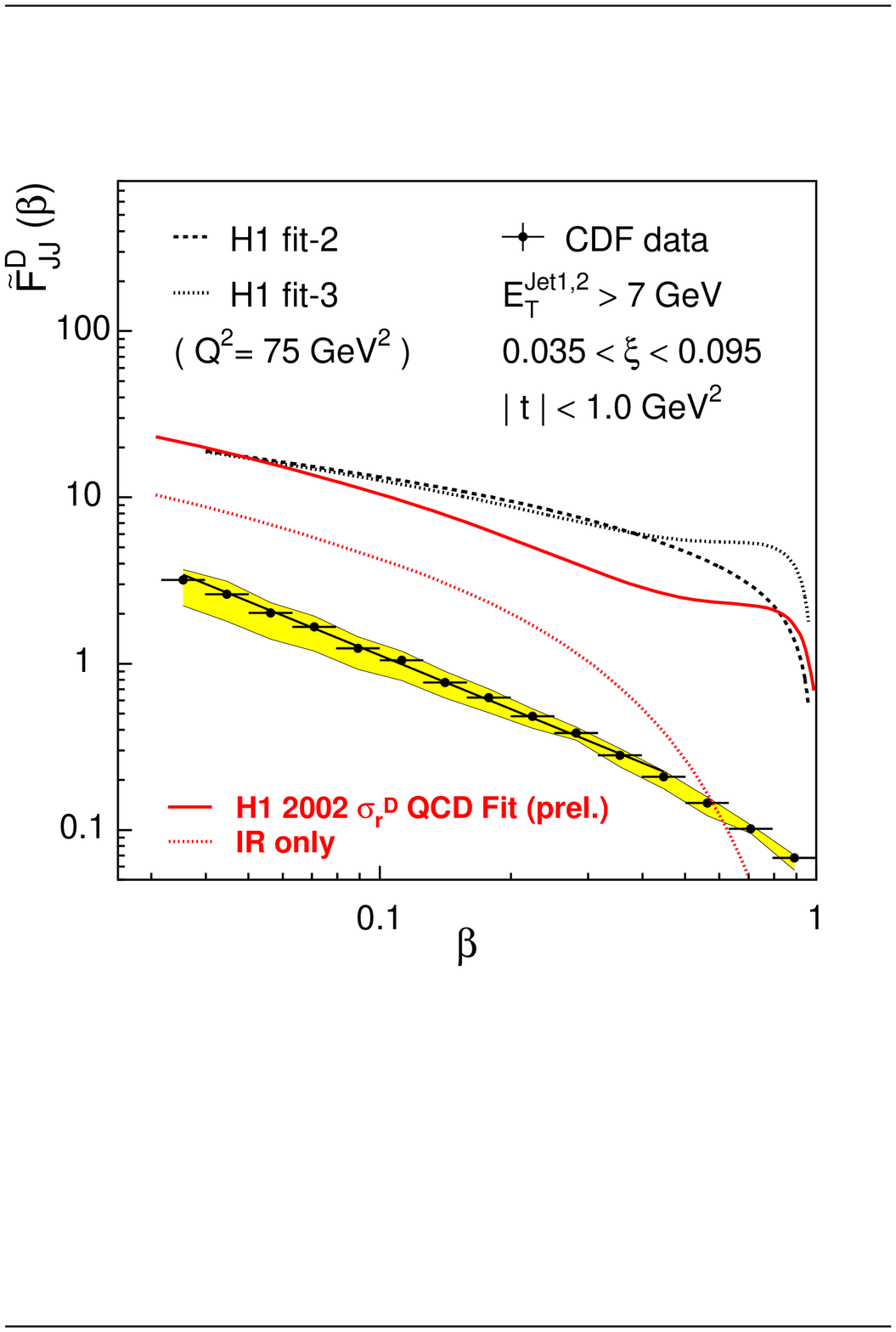}
\end{center}
\caption[]{
Comparison of the diffractive dijet rate to CDF data. (Figure from
\cite{Schilling})}
\label{fig:fdjj}
\end{figure}

However, the problem may not be so bad. With the new H1 partons the
rate for diffractive dijet production at the Tevatron agrees well with
the HERA prediction with a gap survival factor of $0.1$. The Tevatron data
on diffractive dijet production now go beyond the total rate, as can
be seen in Fig. \ref{fig:fdjj} ($\tilde{F}^D_{JJ}$ is a 
ratio of diffractive to non-diffractive cross-sections and
$\beta$ is the momentum fraction of the parton coming from the pomeron
which is involved in the hard subprocess). The solid red
curve shows the prediction based on the latest H1 partons -- it agrees
in shape with the CDF data, especially once one realises that the large 
$\beta$ region is the large $z$ region in Fig. \ref{fig:h1partons} and that
the dijet process is gluon dominated. So much for dijet production in 
single diffraction. What about the double pomeron exchange process
(producing central dijets), which
is out by two orders of magnitude according to Table \ref{tab:teva}?

Remarkably even here things seem not to be too bad. 
Analysis shows that there are large hadronisation
corrections to the parton level results of Table \ref{tab:teva} which
account for a suppression of the theory by a factor of 4 \cite{Appleby}. 
This suppression
comes about because CDF used a small cone $R=0.7$ to define their central
jets, simultaneously with a low $E_T$ cut of 7 GeV. At such low scales the
jets are broad and one loses about 2 GeV per jet. In
\cite{Appleby}, a HERWIG monte carlo simulation of the parton showering and
hadronisation \cite{pomwig} was included thereby allowing us to estimate the
size of the corrections. In addition, the overall rate is sensitive to the
pomeron intercept. A higher intercept leads to a lower cross-section at the
Tevatron if the normalisation is fixed at HERA since HERA probes lower 
values of $\xpom$. The authors of \cite{Whitmore} used a 
soft pomeron intercept which we
now know to be inappropriate; this gains another factor of 2. The remaining
difference is down to the parton densities in the pomeron. Table 
\ref{tab:rob} summarises the results of \cite{Appleby} where the bottom line
shows a difference of a factor 10 between theory and experiment which is in
accord with gap survival estimates. Note that there is good reason to expect
significant ``contamination'' from secondary exchanges ($\RO$) in the region of
the Tevatron data.
 
\begin{table}
\caption{Rates for double pomeron exchange: comparison of theory and
experiment. Theory calculations performed using H1 fit 3 (LO) partons
and the default secondary exchange of \cite{pomwig}}
\begin{center}
\renewcommand{\arraystretch}{1.4}
\setlength\tabcolsep{5pt}
\begin{tabular}{lcrr}
\hline\noalign{\smallskip}
Regge exchange  & Parton Level (nb) & Hadron Level (nb) \\
\noalign{\smallskip}
\hline
\noalign{\smallskip}
$\PO$ & 1175 & 339 \\
$\RO$ & 241 & 58  \\
$\PO+\RO$ & 1416 & 397 \\
\hline
CDF data & & $43.6 \pm 4.4 \pm 21.6$ \\
\hline
\end{tabular}
\end{center}
\label{tab:rob}
\end{table}

\subsection{Central higgs production}
\begin{figure}
\begin{center}
\includegraphics[width=.3\textwidth]{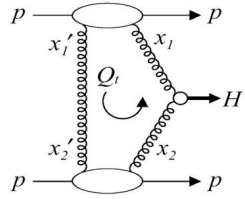}
\end{center}
\caption[]{Exclusive higgs production. (Figure from \cite{KMS})}
\label{fig:higgs}
\end{figure}

It has been suggested that the double pomeron exchange process could be
utilized at the LHC to produce new particles, in particular the higgs
boson \cite{Higgs,KMS}. 
If the higgs is produced exclusively, as shown in Fig. \ref{fig:higgs},
then one could reconstruct its mass quite accurately (to within 1 GeV 
\cite{deRoeck}) by tagging the outgoing protons. Moreover, the exclusive
nature of the central system leads to a significant suppression of QCD
backgrounds, so that one could utilise the $b \bar{b}$ decay of the higgs
with a much better $S/B$ compared to the non-diffractive production
mechanism. Khoze, Martin and Ryskin (KMS) calculate the rate for the diagram in
Fig. \ref{fig:higgs} and find 3~fb for a 115 GeV higgs decaying to
$b \bar{b}$ at the LHC, which should be sufficiently large to permit a 
good measurement. It is possible to test the reliability of this estimate
by performing the corresponding calculation for exclusive central dijet
production (i.e. replace the higgs by a dijet pair). KMS predicted a rate
of around 1~nb at the Tevatron, which is to be compared to the CDF upper
limit of 4~nb \cite{CDF:DPE}, i.e. the process has not been seen in Run I 
data. However,
with the increase in data from Run II, and the fact that D0 now has roman
pot detectors in both forward and backward directions, it should be possible
to check the KMS calculation.  

\section{Dipole models}
\begin{figure}
\begin{center}
\includegraphics[width=.7\textwidth]{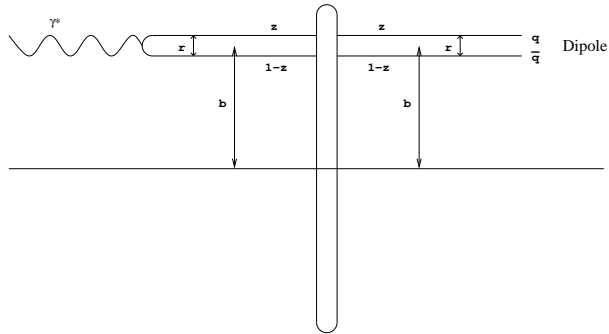}
\end{center}
\caption[]{Diffraction of a colour dipole}
\label{fig:dipoles}
\end{figure}

Complementary to the regge picture of diffraction in photon induced
reactions is the dipole picture, to which we now turn our attention.
In the proton rest frame, the incoming photon converts into a $q \bar{q}$
pair a long distance upstream of the proton. In the diffractive limit,
the quarks are highly energetic and travel along straight lines through
the proton, picking up a non-Abelian phase factor, before eventually
forming the diffracted system $X$ way downstream of the proton, see
Fig. \ref{fig:dipoles}. Consequently, one can write down expressions for
a variety of diffractive process, e.g. for the total $\gamma p$ 
cross-section at high-energy we only need the imaginary part of the forward
elastic scattering amplitude, i.e.
\begin{equation}
\sigma_{\gamma p}^{T,L} = \int dz \, d^2 {\bf r} \, |\psi^{T,L}(z,r)|^2
\sigma(s,r,z)
\end{equation}
where $\sigma$ is the cross-section for scattering a colour dipole of
transverse size $r$ and energy fraction $z$ off a proton. It is universal
in that the same dipole cross-section should appear in other processes, 
such as diffractive vector meson production, 
where one simply replaces the outgoing photon wavefunction with the
meson wavefunction \cite{Nikolaev}. There is clearly a lot of 
physics in the dipole cross-section, including the QCD evolution 
of the original dipole. Apart from the total $\gamma p$ cross-section
(and hence the structure functions, $F_2$, $F_L$ and $F_2^c$) and vector
meson production, the dipole formalism has been used to compare to data on
inclusive diffractive DIS ($F_2^{D(3)}$), deeply virtual Compton scattering
and shadowing off nuclei.

\begin{figure}
\begin{center}
\includegraphics[width=.6\textwidth]{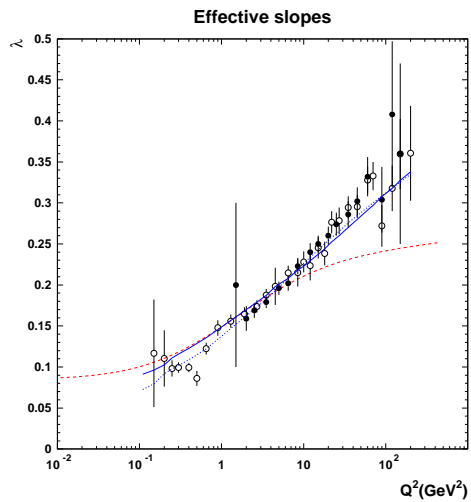}
\end{center}
\caption[]{The effective slope of the low $x$ structure function. Solid
curve is the new dipole model, whilst the dashed curve is that from the
original Golec-Biernat \& W\"usthoff model. (Figure from \cite{Bartels})}
\label{fig:gbwslope}
\end{figure}

In the dipole model of Golec-Biernat \& W\"usthoff \cite{GBW}, 
the dipole cross-section was parameterized as
\begin{equation}
\sigma = \sigma_0 \left\{ 1 - \exp \left( - \frac{r^2}{4 R_0(x)^2} \right)
\right\}
\end{equation}
where
$$ R_0(x) = \frac{1}{{\rm GeV}^2} \left( \frac{x}{x_0}\right)^\lambda $$
and
$$ x = \frac{Q^2+4 m_q^2}{W^2}. $$
$R_0(x)$ is called the saturation radius, since for larger $r$ the 
cross-section flattens off. Since 
the saturation radius moves to smaller $r$ as $x$ decreases this
model naturally tames the powerlike behaviour of the total cross-section
as one moves to smaller $x$. For small enough $r$, the cross-section goes
like $r^2$ which generates Bjorken scaling. The striking agreement
of the model with data $F_2$ and on $F_2^{D(3)}$ originally led to the 
idea that HERA was already probing the non-linear dynamics of saturation.
Subsequent, more detailed studies using the latest data, have
revealed that it is not possible to fit the data with a pure power, 
i.e. $\lambda$, which gets tamed by saturation effects. It is necessary to
replace the exponent with the gluon density \cite{Bartels}, i.e. 
\begin{equation}
\sigma = \sigma_0 \left\{ 1 - \exp \left( - 
\frac{ \pi^2 r^2 \alpha_s xg(x)}{3 \sigma_0} \right)
\right\}.
\end{equation}   
The gluon density itself becomes less steep as $Q^2$ falls and so the
exponentation is less important. This means that the regime of large 
corrections arising from the non-linear dynamics is pushed beyond the HERA 
region, i.e. to smaller $x$. The need to go beyond the original model of
Golec-Biernat \& W\"usthoff is illustrated in Fig. \ref{fig:gbwslope} where
the effective slope ($\sim x^{-\lambda}$) is shown as a function of $Q^2$.
In the original model, this slope asymptotes to just below 0.3 and deviations
at lower $Q^2$ are wholly attributed to saturation dynamics. In the new
model, there is no flattening at high $Q^2$. 

A quite different approach can be found in \cite{FKS}. Here the dipole
cross-section is written as a sum of two terms each of which can be thought
of as arising from a pure regge pole (of intercepts 1.06 and 1.4 
respectively). This ``two pomeron'' model has no saturation dynamics at all. 
The reduction of the effective slope at low
$Q^2$ arises because of the dominance of the pomeron with lower intercept
in that region (the reverse occuring at high $Q^2$). This model is also
able to describe the available data, including the diffractive structure
function $F_2^{D(3)}$ \cite{FKS:f2d3} and deeply virtual Compton scattering 
\cite{MSS:dvcs}.

Before leaving dipoles, I should say a few things about the latest theoretical
progress in the physics of saturation. The use of non-linear perturbative 
QCD dynamics to control the growth of low-$x$ cross-sections has a long
history, dating back to the ``GLR equation'' of the early 1980's \cite{GLR}.
More recently, the Balitsky-Kovchegov equation has been developed to 
describe the non-linear evolution of the S-matrix for scattering a colour
dipole off a hadronic target \cite{BK}. Underpinning all of this is the
colour-glass-dynamics of \cite{CGD}. Formulated as an effective field
theory (analogous to that of glasses in condensed matter physics), the
colour glass dynamics describes the quantum evolution of soft gluons in
a classical background colour field. It reduces to the BFKL equation in
the approximation of a dilute background, and to the Balitsky-Kovchegov
equation in the large $N_c$ limit.

\section{Rapidity gaps at high-$t$}
So far, all the processes we have looked at are close to $t=0$, i.e. the
outgoing proton(s) do not receive a large transverse momentum. Let us now
focus on the case the $-t \gg \Lambda^2$. In this case, the incoming
proton will typically be shattered. It is thought that the largeness of the 
momentum transfer will allow us to utilise QCD perturbation theory and
hence to test the relevance of BFKL dynamics in these processes. 

\subsection{Vector mesons}

\begin{figure}
\begin{center}
\includegraphics[width=.45\textwidth]{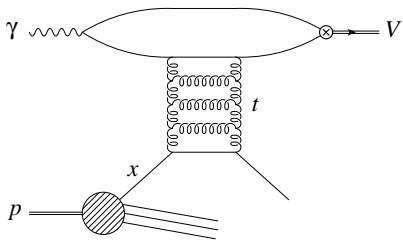}
\end{center}
\caption[]{High $p_T$ vector meson production. (Figure from \cite{EMP})}
\label{fig:mesons}
\end{figure}

In Fig. \ref{fig:mesons} we show the diffractive production of a high
$p_T$ vector meson, $V$. The large momentum transfer across the gap, 
$-t \approx p_T^2$, almost always breaks up the proton and leads to a
jet in the resultant debris. The non-perturbative dynamics is factorised
into either the proton parton density functions or into the meson lightcone
wavefunction, and QCD can be used to compute the dynamics of the exchange.
The fact that $x W^2 \gg -t$ ensures that there is still a large rapidity
gap between the proton dissociation products and the vector meson.

In leading order, BFKL predicts that the hard subprocess cross-section
for scattering off quarks (gluons differ only by a colour factor) 
should go like \cite{Ryskin}
\begin{equation}
\frac{d \sigma(\gamma q \to Vq)}{dt} \sim 
\frac{\alpha_s^4}{t^4} \frac{e^{8z \ln 2}}{z^{3/2}} \exp \left( -
\frac{\ln^2 \tau}{112 z \zeta(3)} \right),
\end{equation}
where
$$ \tau = \frac{-t}{Q^2+m_V^2} ~~ {\rm and} ~~~
z=\frac{3 \alpha_s}{2 \pi} \ln \frac{x W^2}{s_0}. $$

\begin{figure}
\begin{center}
\includegraphics[width=.4\textwidth,angle=-90]{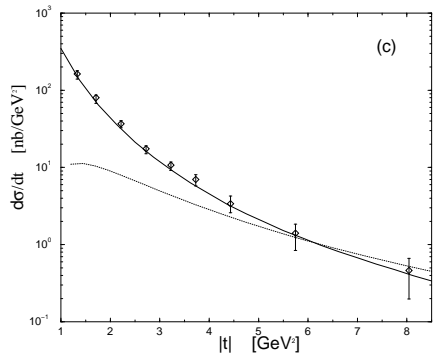}
\end{center}
\caption[]{Comparison of data \cite{ZEUS:mesons} 
with theory for high $p_T$ $\rho$ production.
The solid line is LO BFKL and the dotted line two-gluon exchange. (Figure
from \cite{Gavin})}
\label{fig:rho}
\end{figure}

Comparison of the leading order BFKL calculation with the ZEUS data is
shown in Fig. \ref{fig:rho} \cite{Gavin}. The calculation is a fit to the
data with $s_0 = \beta m_V^2 - \gamma t$ and $\alpha_s$ being the free
parameters. The fit shown corresponds to $\beta = 0$, $\gamma = 1$ and
$\alpha_s = 0.20$. Good agreement is also found (with the same parameters)
with the data on the $\phi$ and $J/\psi$ mesons \cite{Gavin}. 
Note that it is not
possible to get good agreement in the approximation that only two-gluons
are exchanged between the diquark system and the struck parton.

The above curves were computed assuming a very simple form for the meson
wavefunction. In particular, it is assumed that the quarks share equally
the momentum of the meson. Relativistic corrections to this simple
approximation have been considered \cite{Ivanov} 
and do not appear to spoil the good
agreement \cite{EMP}. Inclusion of relativistic corrections also allows one to
quantify the degree to which the helicity of the meson differs from that
of the photon, and comparison with data is once again encouraging \cite{EMP}.

In the future, data will become available on high-$t$ photon production. 
This is something to look forward to, since it avoids the uncertainty
associated with the production of the vector meson.

\subsection{Gaps between jets}

\begin{figure}
\begin{center}
\includegraphics[width=.45\textwidth]{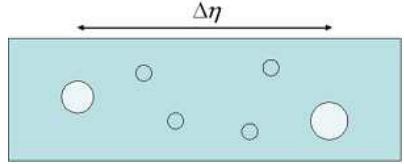}
\end{center}
\caption[]{Dijet production with a rapidity gap}
\label{fig:lego}
\end{figure}

As well as vector meson production, one can look for rapidity gaps between
jets in photon-hadron and hadron-hadron collisions \cite{Tang}. 
The typical final 
state topology is shown in Fig. \ref{fig:lego}, where two jets are produced
far apart in rapidity and there is a gap between the jets. Early measurements
at HERA \cite{gbj:zeus} and the Tevatron \cite{gbj:cdf,gbj:d0} 
have been compared to theory and leading order
BFKL does fine \cite{CFL,Enberg}. 
However, conclusive statements are hard to make either because the 
gap is not large enough (i.e. the excess over non-BFKL QCD is small)
or hadronisation corrections are large \cite{CFL}.

\begin{figure}
\begin{center}
\includegraphics[width=0.95\textwidth]{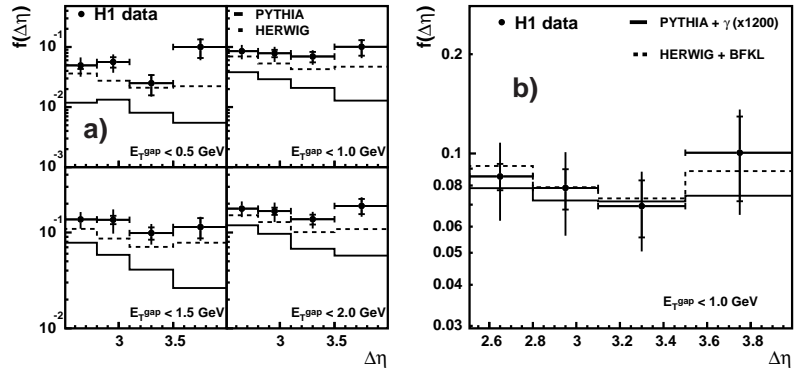}
\end{center}
\caption[]{Recent H1 results on gaps between jets shown as a function of
rapidity gap and transverse energy between the jets. 
(Figure from \cite{Wyatt})}
\label{fig:H1}
\end{figure}

There has been significant recent progress, both experimentally and
theoretically in this area which has to some extent shifted interest away
from BFKL dynamics and into the domain of jet energy flows. H1 has focussed
on the definition of the gap. They use the $k_T$ cluster algorithm to put all
hadrons into jets, after which they select the two highest $p_T$ jets. The
summed transverse energy between these two jets ($E_T^{{\rm gap}}$) is then
used to define a gap. As can be seen in Fig. \ref{fig:H1}, for 
low $E_T^{{\rm gap}}$, one can really speak of
a rapidity gap and a very clear excess is seen in the data over the standard
Monte Carlos, whilst at larger
values of $E_T^{{\rm gap}}$ the enhancement is less pronounced \cite{Wyatt}. 
By defining
their gaps this way, H1 has reduced its sensitivity to soft gluon radiation by
effectively cleaning up the edge of the gap in a way which makes possible
direct comparison with future theoretical calculations. 

On the theoretical side, Dasgupta \& Salam have recently pointed out that
there is a previously unconsidered mechanism which ought to be considered
when considered interjet energy flows \cite{DS}. 
In particular, they have discovered
a class of ``non-global'' logarithms which ought to be summed at the single
logarithm level. 

\section{Summary}
The key conclusions of this talk can be summarised as follows:
\begin{itemize}
\item{Regge factorisation and QCD evolution work well to describe 
diffractive deep inelastic scattering at HERA}
\item{HERA partons may well be useful at the Tevatron provided one
accounts for gap survival at a level of around $10\%$. Tevatron
measurements at lower $\xpom$ would help.}
\item{There is a need to measure exclusive central dijet production
at Tevatron Run II in order to constrain diffractive higgs cross-sections
at the LHC.}
\item{Position on saturation is not clear. Ideally one would like to go
to lower $x$.}
\item{High-$t$ vector meson production is well described by leading order
BFKL.}
\end{itemize}

\section*{Acknowledgements}
Special thanks to Brian Cox, Rikard Enberg and Paul Newman for their
help in preparing this talk. Thanks also to the conference organisers
for their invitation to participate.

\end{document}